\documentclass{aastex}
\usepackage{graphicx}
\usepackage{emulateapj5}

\submitted{Accepted by AJ - 2002/11/04}

\shortauthors{Hopkins et al.}
\shorttitle{PDS: 1.4\,GHz $\mu$Jy catalogue}

\begin{document}

\title{The Phoenix Deep Survey: The 1.4\,GHz microJansky catalogue}

\author{A. M. Hopkins\altaffilmark{1,2}, J. Afonso\altaffilmark{3,4},
        B. Chan\altaffilmark{5}, L. E. Cram\altaffilmark{6},
        A. Georgakakis\altaffilmark{7}, B. Mobasher\altaffilmark{8}
}

\affil{\begin{enumerate}
\item Department of Physics and Astronomy, University of Pittsburgh, 
  3941 O'Hara Street, Pittsburgh, PA 15260, USA
\item Hubble Fellow; email ahopkins@phyast.pitt.edu
\item Centro de Astronomia e Astrof\'{\i}sica da Universidade de Lisboa,
   Observat\'orio Astron\'omico de Lisboa, Tapada da Ajuda,
   1349-018 Lisboa, Portugal
\item Blackett Laboratory, Imperial College, Prince Consort Road,
   London SW7 2BW, UK
\item School of Physics, University of Sydney,
  NSW 2006, Australia
\item Australian Research Council, GPO Box 9880, Canberra ACT 2601,
  Australia
\item National Athens Observatory, Institute of Astronomy \& Astrophysics,
 I.Metaxa \& Vas.Pavlou str., Athens 15236, Greece
\item Space Telescope Science Institute, 3700 San Martin Drive,
 Baltimore, MD 21218, USA
\end{enumerate}
}

\begin{abstract}
The initial Phoenix Deep Survey (PDS) observations with the Australia
Telescope Compact Array have been supplemented by additional 1.4\,GHz
observations over the past few years. Here we present details of the
construction of a new mosaic image covering an area of 4.56 square degrees,
an investigation of the reliability of the source measurements,
and the 1.4\,GHz source counts for the compiled radio catalogue.
The mosaic achieves a $1\sigma$ rms noise of $12\,\mu$Jy
at its most sensitive, and a homogeneous radio-selected catalogue of
over 2000 sources reaching flux densities as faint as $60\,\mu$Jy has been
compiled. The source parameter measurements are found to be consistent with the
expected uncertainties from the image noise levels and the Gaussian source
fitting procedure. A radio-selected sample avoids the complications of
obscuration associated with optically-selected samples, and by utilising
complementary PDS observations including multicolour optical, near-infrared
and spectroscopic data, this radio catalogue will be used in a detailed
investigation of the evolution in star-formation spanning the redshift
range $0 < z < 1$. The homogeneity of the catalogue ensures a consistent
picture of galaxy evolution can be developed over the full cosmologically
significant redshift range of interest.
The 1.4\,GHz mosaic image and the source catalogue are available on the web at
\verb+http://www.atnf.csiro.au/~ahopkins/phoenix/+
or from the authors by request.
\end{abstract}

\keywords{galaxies: general --- radio continuum: galaxies --- galaxies: starburst --- galaxies: evolution --- surveys}

\section{Introduction}
\label{int}

Exploration of star-formation processes in galaxies through observations
at radio wavelengths has developed rapidly in the past few years
\citep{Hop:98b,Rich:98,Hop:99b,Rich:00,Pran:01,Masc:01,Sad:02,deV:02}.
By avoiding the problems associated with dust obscuration, dominant
at optical wavelengths, radio data provide a valuable tool in understanding
the full picture of star-formation in galaxies. The evolution of these
galaxies, and the role of interactions and mergers in the population, are
still only partially understood.

The Phoenix Deep Survey (PDS) aims to catalogue a large sample of
{\em radio-selected\/} star-forming galaxies spanning the redshift range
$0<z<1$. The PDS includes a 1.4\,GHz survey made using the Australia Telescope
Compact Array (ATCA), and covers a field a little more than $2^{\circ}$
diameter, selected to lie in a region of low optical obscuration and devoid of
bright radio sources \citep{Hop:98b}. To clarify our nomenclature, we use the
acronym PDS to designate the survey as a whole, which includes multicolour
optical photometry and spectroscopy as well as the radio imaging. The full
1.4\,GHz ATCA mosaic itself, over which all the complementary multiwavelength
observations have been conducted, is referred to as the Phoenix Deep Field
(PDF). The PDF provides a large, homogeneous sample of
1.4\,GHz sources that, through the complementary multiwavelength observations
of the PDS, is being used to investigate star forming galaxies
in the faint radio population. Existing analyses of the PDS
\citep{Afo:02,Afo:01a,Afo:01b,Afo:01c,Afo:99,Age:99c,Age:00b,Age:99a,Age:99b,Hop:00,Hop:99a,Hop:99b,Hop:98b,Hop:98a,Mob:01,Mob:99}
have described the initial survey and catalogues, and have initiated
investigations into numerous aspects of the sample, including some
implications for star-formation processes in galaxies.

Recent ATCA observations have improved the sensitivity over a larger
area within the PDF, allowing a larger radio catalogue to be compiled
and increasing the number of radio sources with optical counterparts.
The new mosaic image is slightly larger than the original mosaic
\citep{Hop:98b}, and the catalogues constructed from it contain over 2000
identified radio sources. This corresponds to roughly a thirty percent
increase over the number of sources available from the complete catalogue
used in the analysis of \citet{Hop:99b} and twice the number from the
original catalogue described by \citet{Hop:98b}. The additional source
detections are a result of expanding the region over which we achieve the
level of greatest sensitivity, comparable to that initially reported by
\citet{Hop:99b}.
In \S~\ref{obs} of this paper we present the 1.4\,GHz ATCA observations
and summarise the image processing; \S~\ref{sources} describes the radio
source measurements and their reliability; \S~\ref{catalogues} explains
the compilation of the radio source catalogues; \S~\ref{scounts}
describes the construction of the 1.4\,GHz source counts;
\S~\ref{results} presents the source count results; \S~\ref{disc}
provides a discussion of the similarities to and discrepancies with other
source count estimates;
and \S~\ref{summ} summarises our main results and conclusions.

\section{Observations}
\label{obs}

The PDF \citep{Hop:99b,Hop:98b,Hop:98a} covers a high-latitude region
of low optical obscuration and devoid of bright
radio sources. ATCA 1.4\,GHz observations were made in 1994, 1997, 1999,
2000 and 2001, in the 6A, 6B and 6C array configurations, accumulating
a total of 523 hours of observing time. The initial 1994 ATCA observations
\citep{Hop:98b,Hop:98a} consisted of 30 pointings on a hexagonal
tesselation, resulting in a $2^{\circ}$ diameter field centered on
RA(2000)=$01^{\rm h}14^{\rm m}12\fs16$ Dec(2000)=$-45^{\circ}44'8\farcs0$
with roughly uniform sensitivity of about $60\,\mu$Jy rms. This survey was
supplemented from 1997 to 2001 by extensive observations of a further 19
pointings situated on a more finely spaced hexagonal grid, centered on
RA(2000)=$01^{\rm h}11^{\rm m}13\fs0$ Dec(2000)=$-45^{\circ}45'00''$.
The locations of all pointing centers are given in Table~\ref{points}.

Processing of the data was performed using the {\sc miriad} (Multichannel
Image Reconstruction, Image Analysis and Display) software package,
following the steps detailed in \citet{Hop:98b} and \citet{Hop:99b}.
Some 1994 data were reprocessed to improve both the flagging of data
affected by strong interference and the self-calibration of the brightest
source in the field (a 115\,mJy source towards the eastern edge of the
mosaic). In the 1997-2001 observations, some of the westernmost pointings
showed side-lobe artifacts from a bright out-of-field source. For each
affected pointing, these artifacts were removed as follows. First the object
was imaged by applying an offset to the image center using {\sc miriad}'s
{\sc invert} task, revealing a double component source. The {\sc mfclean}
task was used to create a model for the source, and this model was subtracted
from the flagged data using the task {\sc uvmodel}. Subsequent imaging of
the affected fields (after {\sc mfclean}ing) showed no residual
side-lobe artifacts from this source.

\begin{figure*}
\centerline{\includegraphics[width=14cm]{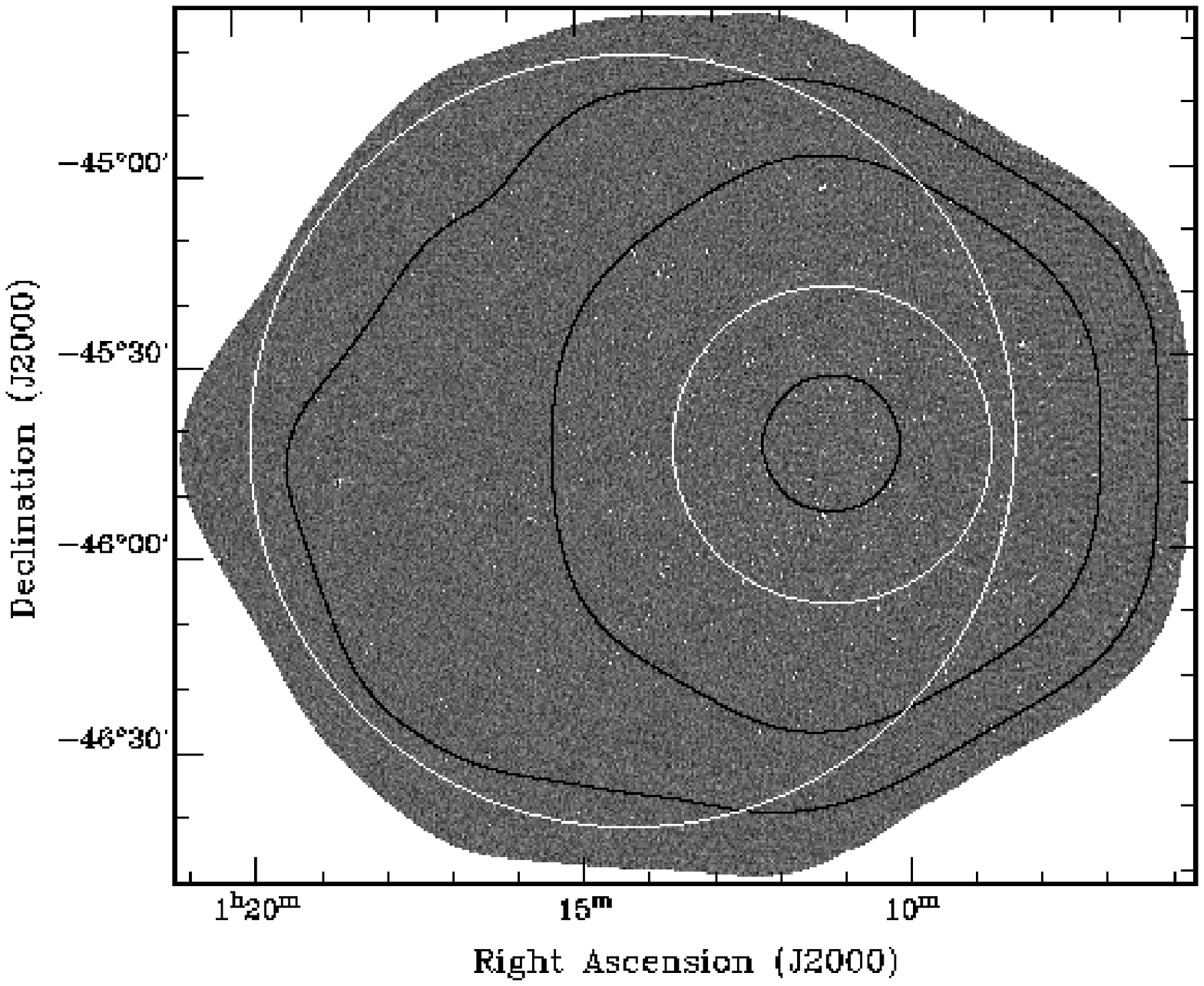}}
\caption{Complete Phoenix Deep Field mosaic. The greyscale image has been
trimmed at the $250\,\mu$Jy rms contour. The large white circle shows the
2$^{\circ}$ diameter region of the original PDF \protect\citep{Hop:98b}.
The small white circle corresponds to the $25\,\mu$Jy rms contour as
measured in an earlier mosaic, which delimited the $50'$ diameter deep
subregion described in \protect\citet{Hop:99b}. (The $25\,\mu$Jy rms contour
of the current mosaic is not shown here, but covers a larger area, see
Figure~\ref{rmscont}.) The black contours mark the theoretical
10, 30 and $90\,\mu$Jy rms levels of this mosaic, emphasising the high
sensitivity over a region roughly $2^{\circ}$ in diameter.
This image has been normalised by subtracting locally determined means and
dividing by locally determined rms noise values, to emphasise the sources
rather than the noise characteristics. Note the uniformity of the background
once the varying noise-level has been accounted for, and the apparent
predominance of sources in the regions corresponding to the greatest
sensitivity of the survey.
  \label{mosaiccirc}}
\end{figure*}

\begin{figure*}
\centerline{\includegraphics[angle=-90,width=12.0cm]{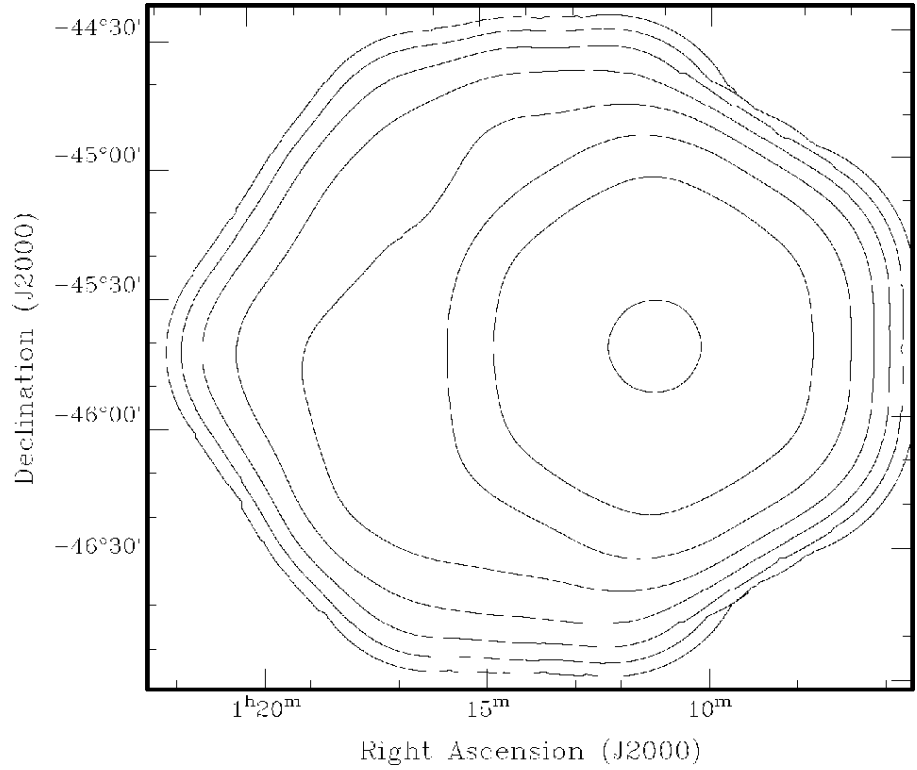}}
\caption{Contours showing the theoretical rms noise level over the
full 1.4\,GHz mosaic, ranging from $10\,\mu$Jy (inner circle) to 1.28\,mJy,
in steps of factors of two.
  \label{rmscont}}
\end{figure*}

\begin{figure*}
\centerline{\includegraphics[width=12.0cm]{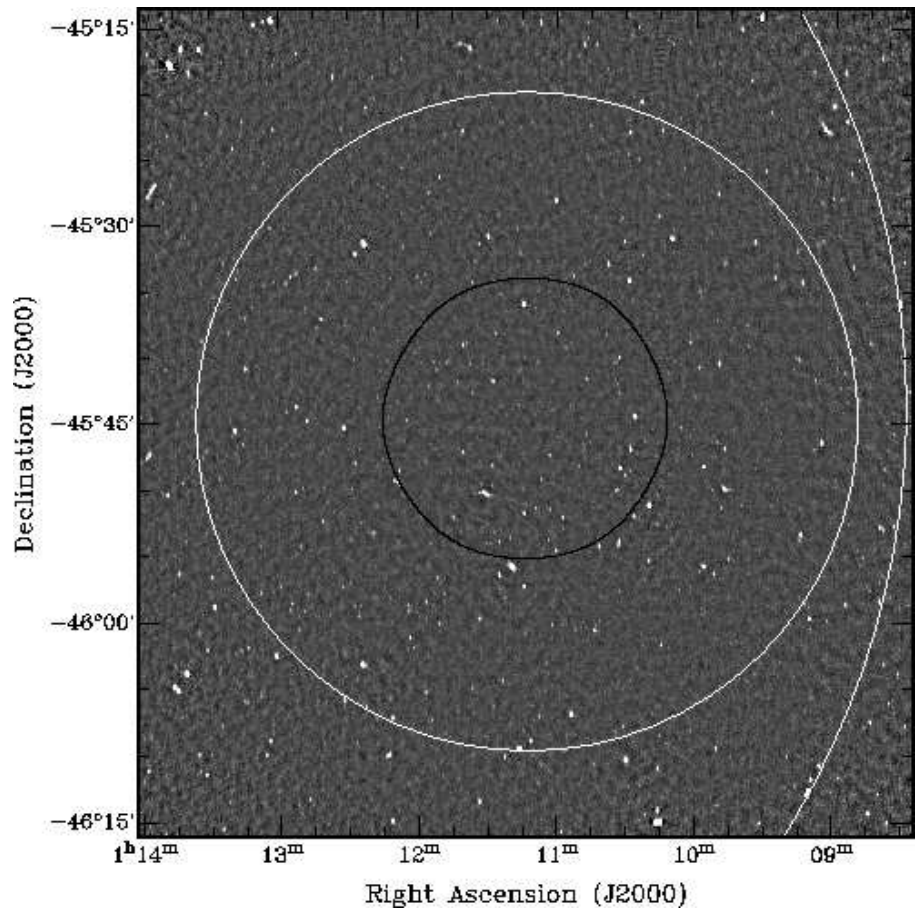}}
\caption{A magnified view of a $63'\times63'$ region centered on
the most sensitive portion of the new PDF mosaic, showing a selection
of sources in more detail. Circles and contour as in Figure~\ref{mosaiccirc}.
  \label{pdf1}}
\end{figure*}

After using the task {\sc mfclean} to construct clean-component models for
each pointing, the task {\sc restor} was used to make a clean image for
each field, specifying a fixed restoring beam size of $6''\times12''$
(in the RA$\times$Dec sense) for all pointings. This was done in
anticipation of the mosaicing step, to ensure that common sources
in overlapping pointings were not represented by differently sized
or oriented intensity distributions. The final mosaic was constructed from
all 49 pointings using the task {\sc linmos}. The mosaic was then trimmed
to remove the highest noise regions at the edges by masking out regions
with an rms noise level greater than 0.25\,mJy. This resulted in an image
of $4781\times4111$ pixels (with a pixel scale of $2''$ pixel$^{-1}$),
shown in Figure~\ref{mosaiccirc}. Figure~\ref{rmscont} shows contours
corresponding to the theoretical rms noise level for the mosaic (but see
also Figure~\ref{areavsflx} and the discussion in \S~\ref{scounts} below),
and a magnified view of the most sensitive portion of the mosaic is shown in
Figure~\ref{pdf1}. The trimmed PDF mosaic image now covers an area of 4.56
square degrees and reaches to a measured level of $12\,\mu$Jy rms noise in the
most sensitive regions. With the addition of the more sensitive 1997-2001
data, the noise characteristics over the field are no longer uniform,
as was the goal of the original 1994 project \citep{Hop:98a}. This is
not so much a drawback as might be supposed however, as will be seen in
the next section describing source-detection. Indeed, the combination of
the deeper imaging with a shallower wide-field presents an opportunity
to simultaneously explore properties of both the faintest radio sources
and the brighter ones, from levels of about $50\,\mu$Jy to almost 50\,mJy,
although the sampling of the brightest sources will clearly still be sparser
than the faintest.

Source confusion is not yet a concern for the survey, even at the flux
density levels reached. From the number of sources detected and the area
sampled (see \S~\ref{catalogues} below), even in the region of highest source
surface density close to the flux density limits, there are still about
140 independent beam areas per source. This is sufficient to ensure that
source confusion should not bias the survey, although observations to deeper
levels at this resolution may start to be affected.

\section{Source detection and measurement}
\label{sources}

The {\sc miriad} task {\sc sfind} \citep{Hop:02} was used for detecting
sources in the PDF mosaic. In an image such as the PDF mosaic with a
non-uniform mean and a continuously varying noise level, source detection by
specifying a uniform threshold over the complete image is clearly not
the correct approach. Calculation of locally determined rms noise levels
over the image is necessary, and the {\sc sfind} task implements
this by dividing the image into small square regions of a user-specified size
within which the mean and rms noise level are estimated. These quantities
are found by fitting a Gaussian to the pixel histogram in each region,
after iterative sigma-clipping to remove any bias from source-pixels. The image
is then ``normalised" by subtracting the mean and dividing by the rms within
each region, resulting in an image where pixel values are effectively
specified in units of $\sigma$, the local rms noise level. This normalised
image can be used for defining a threshold for source-detection,
specifying which pixels to use in measuring the source parameters, which
is then performed using the original image. The ``normalised" PDF mosaic
(constructed using regions of 100$\times$100 pixels for estimating the
noise characteristics) is shown in Figure~\ref{mosaiccirc}.

By normalising the image before performing the thresholding for
the source-detection, the non-uniform noise level of the original
mosaic is no longer an issue when constructing a source catalogue.
Deciding how to choose the threshold is the next important question.
Traditionally, levels such as $5\sigma$ or even $7\sigma$ have been
chosen to ensure a minimum number of falsely detected sources. Such
strict thresholds bring some level of confidence in the detected
sources at the expense of overlooking many real fainter sources.
Lower thresholds, on the other hand, detect many more sources but
with a higher probability that any given source is spurious. One
way of addressing this issue is to use a statistical technique
called the False Discovery Rate (FDR), which assigns a threshold
based on an acceptable rate of false detections \citep{Hop:02,Mil:01}. 
As described in detail by those authors, an arbitrarily chosen significance
threshold such as $5\sigma$ can fix the total number of false detections
in an image, (which will depend for example on the image size and
sampling), independently of the number of real sources detected. The FDR
method, in contrast, controls what is perhaps a more relevant quantity,
the average {\em fraction\/} of false discoveries over the total number
of discoveries \citep{Mil:01}. The FDR method achieves this by adapting
the threshold to the data (image background plus sources) being analysed.
By specifying that, say, 10\% of the detected sources are allowed to
be false detections, a specific threshold would be defined for a given image
that would ensure no more than 10\% of the detections would be false. There
is a further subtlety here deriving from the relation of astronomical sources
to the source-pixels comprising them in an image, since it is actually the
source-pixels whose false discovery rate is governed by the FDR technique.
This is explored in detail by \citet{Hop:02}, who conclude that in general
the false discovery rate specified for the pixels will correspond fairly
closely to that for the astronomical sources of interest. Deviations
of the image background from true Gaussian noise though, as may occur due
to residual imaging artifacts for example, may act to increase the number
of false detections somewhat, although this is true of any thresholding
technique.

\begin{figure*}
\centerline{\includegraphics[angle=-90,width=13cm]{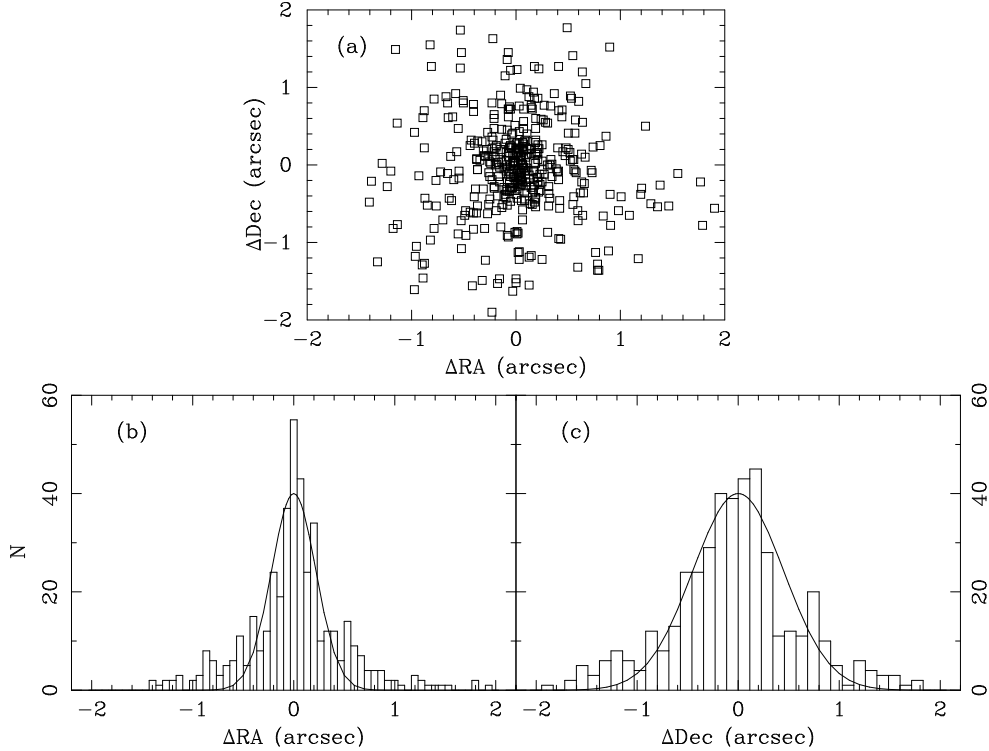}}
\caption{(a) Relative positional offsets in Right Ascension and Declination
for common objects detected independently in overlapping fields. Common
sources were identified as those having positions separated by less than
$2''$, and this is reflected in the limits of the uncertainties shown here.
(b,c) Histograms of relative positional offsets for common objects. The
solid lines are Gaussian fits to the histograms, and indicate the magnitude of
the relative positional uncertainties. For RA the Gaussian shown has
$\sigma=0.22''$, for Dec it has $\sigma=0.45''$. These values are
a result of the fact that the synthesised beam shape is twice as wide in
Dec as in RA. The presence of the outliers above the Gaussian fits
are due to the presence of both intrinsically non-Gaussian sources as well
as falsely-detected pixels that skew the measurements for some objects.
In general the rms positional uncertainties from the {\sc sfind}
measurements should be better than about $0.5''$.
  \label{poserr}}
\end{figure*}

The reliability of the source detection for the PDF using {\sc sfind} 
was investigated by taking advantage of the overlapping nature of
the many pointings used to construct the mosaic. Several overlapping fields
were first corrected for the attenuation of the primary beam sensitivity,
after which source detection was performed independently on each.
The task {\sc sfind} was used for the source detection, specifying a
false discovery rate of $10\%$.
Common sources between each pair of fields were identified by positional
matching, assuming that a positional offset of less than $2''$ indicated
a common source. The measurements of the source parameters (positions
and flux densities) for common sources were compared to ascertain their
reliability. In Figure~\ref{poserr} the positional uncertainties are
shown. The histograms in Figure~\ref{poserr} reflect the fact that
the synthesised beam size is twice as large in Declination as in
Right Ascension, and indicate that the relative rms positional uncertainty
from the {\sc sfind} measurements is less than about $0.5''$.

To investigate the reliability of the flux density measurements, the
uncertainties in the measurements need to first be established. The relative
error in the integrated flux density, $I$, for a source was estimated from
\begin{equation}
\label{ierr}
\frac{\sigma_{I}}{I} = \sqrt{\frac{\mu_{\rm image}^2}{I^2}
 + \frac{\mu_{\rm fit}^2}{I^2}}
\end{equation}
where $\mu_{\rm image}$ is the uncertainty due to the rms noise in the image,
and $\mu_{\rm fit}$ is the uncertainty in the Gaussian fitting. From
\citet{Wind:84}
\begin{equation}
\frac{\mu_{\rm image}}{I} = \sqrt{\frac{\sigma^2}{S^2} + C_f^2 + C_p^2}
\end{equation}
where $\sigma$ is the rms noise at the location of the source, $S$ is the
peak flux density of the source, $C_f$ is the relative error in the
absolute flux calibration and $C_p$ is the relative flux error introduced
due to pointing errors of the individual telescope dishes. The latter
two terms combined are of the order of 1\% ($C_f^2 + C_p^2=0.01^2$), and
this value has been used for subsequent calculations.
The rms noise in the image is correlated over the synthesised beam area,
and from equation~42 of \citet{Con:97} the relative uncertainty from the
fitting is taken to be
\begin{equation}
\frac{\mu_{\rm fit}^2}{I^2} = \frac{\mu_S^2}{S^2}
 + \left(\frac{\theta_B\theta_b}{\theta_M\theta_m}\right)
  \left[\frac{\mu_M^2}{\theta_M^2} + \frac{\mu_m^2}{\theta_m^2} \right]
\end{equation}
where $\theta_B\theta_b$, the product of the synthesised beam major and
minor axis full width at half maxima, is used in place of Condon's
$\theta_N^2$. (Condon assumes a circular Gaussian shape for the smoothing
corresponding to the area over which the noise is correlated, whereas we
have an elliptical Gaussian shape.) The
uncertainties $\mu_S$, $\mu_M$ and $\mu_m$ are those due to the fitting in
the peak flux density $S$, major axis $\theta_M$ and minor axis $\theta_m$
respectively. These are approximated by Condon's equation~21,
$\mu_X^2/X^2 \approx 2/\rho_X^2$, where $\rho_X$,
the signal-to-noise ratio (S/N) of the fit, is parameter dependent, as
given by Condon's equation~41:
\begin{equation}
\rho^2 = \frac{\theta_M\theta_m}{4\theta_B\theta_b}
  \left[1+ \left(\frac{\theta_B}{\theta_M}\right)^2 \right]^{\alpha_M}
  \left[1+ \left(\frac{\theta_b}{\theta_m}\right)^2 \right]^{\alpha_m}
  \frac{S^2}{\sigma^2},
\end{equation}
which we have again modified, using $\theta_B$ and $\theta_b$ in place
of Condon's $\theta_N$ in the appropriate places. The parameter dependence
enters through the exponents $\alpha_M$ and $\alpha_m$, which, again
following \citet{Con:97}, we take to be $\alpha_M=\alpha_m=1.5$ for
calculating $\mu_S^2/S^2$, $\alpha_M=2.5$ and $\alpha_m=0.5$ for
$\mu_M^2/\theta_M^2$, and $\alpha_M=0.5$, $\alpha_m=2.5$ for
$\mu_m^2/\theta_m^2$. For point sources, ($\theta_M=\theta_B$,
$\theta_m=\theta_b$, $I=S$), the total relative uncertainty in the integrated
flux density reduces to
\begin{equation}
\label{pointerrs}
\frac{\sigma_I}{I} = \sqrt{2.5 \frac{\sigma^2}{I^2} + 0.01^2}
\end{equation}
\citep[compare with Equation~9 of][]{Ren:97}.

\begin{figure*}
\centerline{\includegraphics[angle=-90,width=15cm]{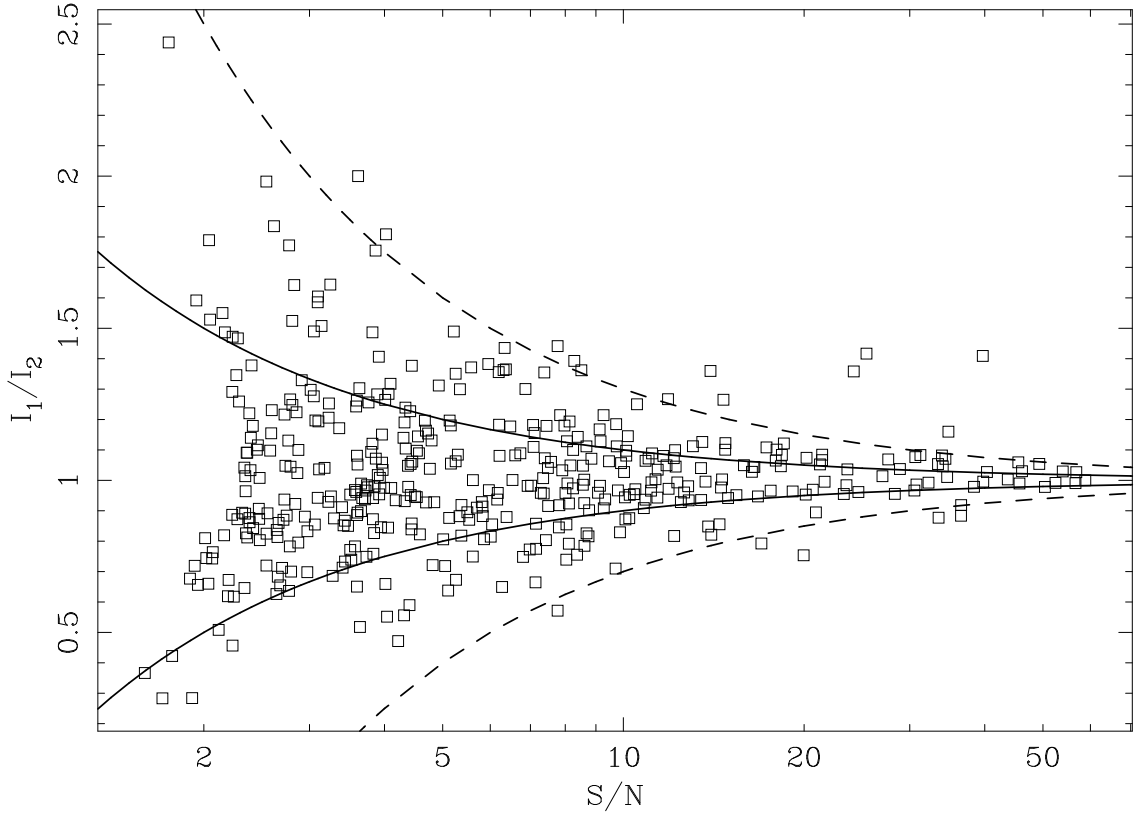}}
\caption{Integrated flux density ratios as a function of the combined S/N for
common sources measured in independent observations. The solid and dashed
lines indicate the $1\sigma$ and $3\sigma$ uncertainties expected from
Equation~\ref{ierr}.
  \label{relflux}}
\end{figure*}

With the value of $\sigma_I/I$ now determined for each source, we can
explore the flux density measurements of common sources detected in
independent observations. For each pair of detections of a common source,
the combined uncertainty is simply the quadrature sum of the individual
relative uncertainties. The S/N for each pair is then taken to be the
inverse of this combined uncertainty. The ratio of integrated flux density
measurements $I_1/I_2$ is shown plotted as a function of the combined S/N
in Figure~\ref{relflux}.
The uncertainties in the flux density ratios (the inverse of the combined
S/N), are shown as the solid and dashed lines, indicating the $1\sigma$
and $3\sigma$ uncertainty levels respectively. There are a few outliers
in this diagram that do not follow the expected uncertainty distribution.
These will be examined further below.

A second flux density measurement comparison was done following
the method used by \citet{Wind:84}. By subtracting unity from the flux
ratios of independent measurements for common objects, then dividing by
the combined rms for the measurements, they aimed to construct a normally
distributed statistic, $A=(I_1/I_2 - 1) / \sqrt{\sigma_1^2+\sigma_2^2}$.
A histogram of this statistic for all
common objects in a pair of overlapping fields can be compared with
a Gaussian of zero mean and unit rms to establish the reliability of
the flux density measurements \citep[their Figures~4e and 4f]{Wind:84}.
This particular statistic, though, is not symmetrical about zero when the
flux density measurements are switched, suggesting that there may be some
skew in the distribution rather than truly being normally distributed. To
avoid this problem, the following statistic was used:
\begin{equation}
\label{a1}
A_1 = \frac{(I_1 - I_2)}{\sqrt{\sigma_1^2 + \sigma_2^2}\ (I_1+I_2)/2}.
\end{equation}
A related statistic, giving very similar results is:
\begin{equation}
A_2 = \frac{(I_1 - I_2)}{(I_1 \times I_2)\sqrt{(\sigma_1/I_1)^2 +
 (\sigma_2/I_2)^2}}.
\end{equation}

\begin{figure*}
\centerline{\includegraphics[angle=-90,width=15cm]{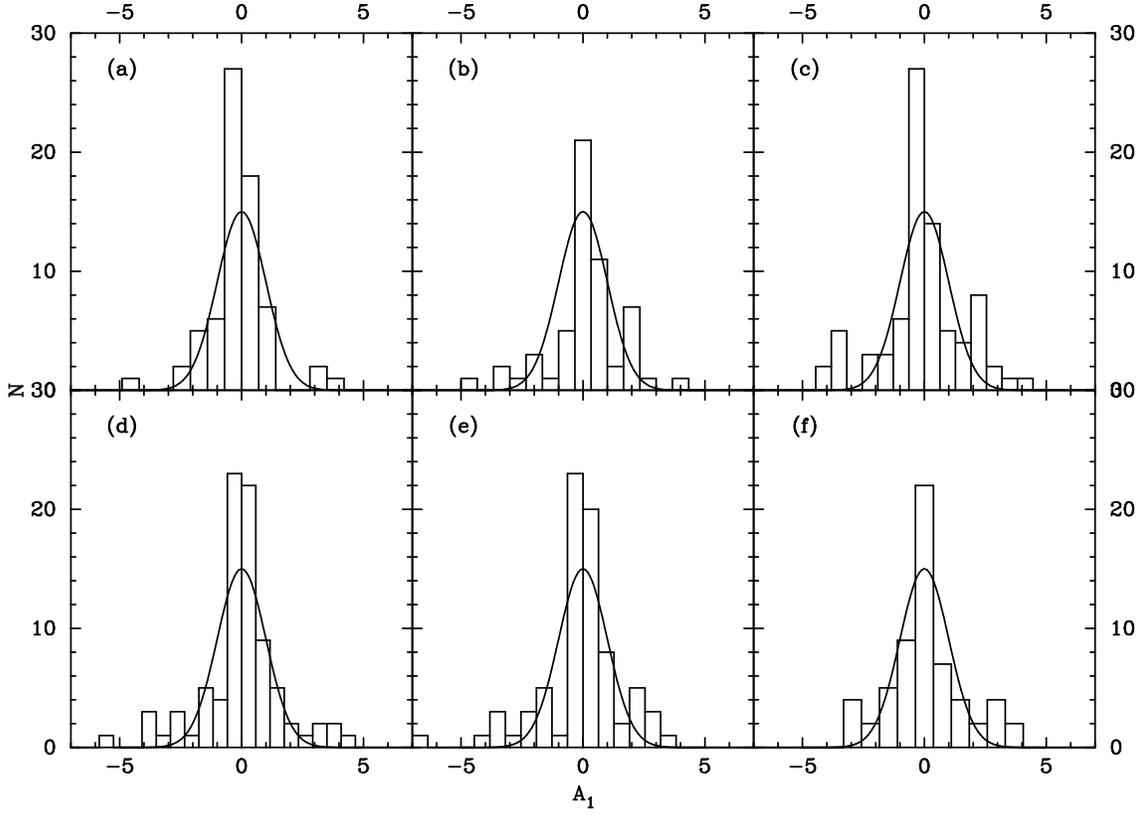}}
\caption{Histograms showing normalised flux ratios, $A_1$, given by
Equation~\ref{a1}. Each histogram contains data for common sources
between a given pair of fields, for 6 pairs examined. The Gaussians
are not fits to the data, but are shown to guide the eye. They have
zero mean, unit rms and peak amplitude of 15. The histogram bin widths
were chosen to produce histograms of an appropriate amplitude,
for convenience of comparison. The majority of sources seem to be
consistent with the expected uncertainties, and the small number of
outliers in each field are consistent with the FDR threshold used.
  \label{fluxcomp}}
\end{figure*}

Histograms of $A_1$ are shown in Figure~\ref{fluxcomp} for six pairs of
fields, the central field in the hexagonal tesselation of the 1997-2001
observations matched with each of the surrounding six fields (see
Table~\ref{points}). The majority of sources seem to be consistent with
the expected uncertainties, and the few outliers are consistent with the
FDR threshold used. Less than 10\% of the sources detected are seen as
outliers here, as expected from the FDR threshold defined by allowing
up to 10\% false detections. The outliers found by \citet{Wind:84}
were variable or extended sources whose flux density comparisons would
not be expected to follow such a distribution. As the fields compared
here were all observed at the same time, variable sources should not
make a significant contribution. Extended, intrinsically non-Gaussian
sources may account for a small fraction of these outliers, but
the majority of the detected sources are not significantly extended,
and the Gaussian model should be reasonable. Visual inspection of many of
the objects responsible for the outliers in Figures~\ref{relflux} and
\ref{fluxcomp} suggest that almost all of them are explained by variations
in the Gaussian fits due to the presence of falsely-detected pixels included
with the source-pixels during the source measurement.

\begin{figure*}
\centerline{\includegraphics[angle=-90,width=14cm]{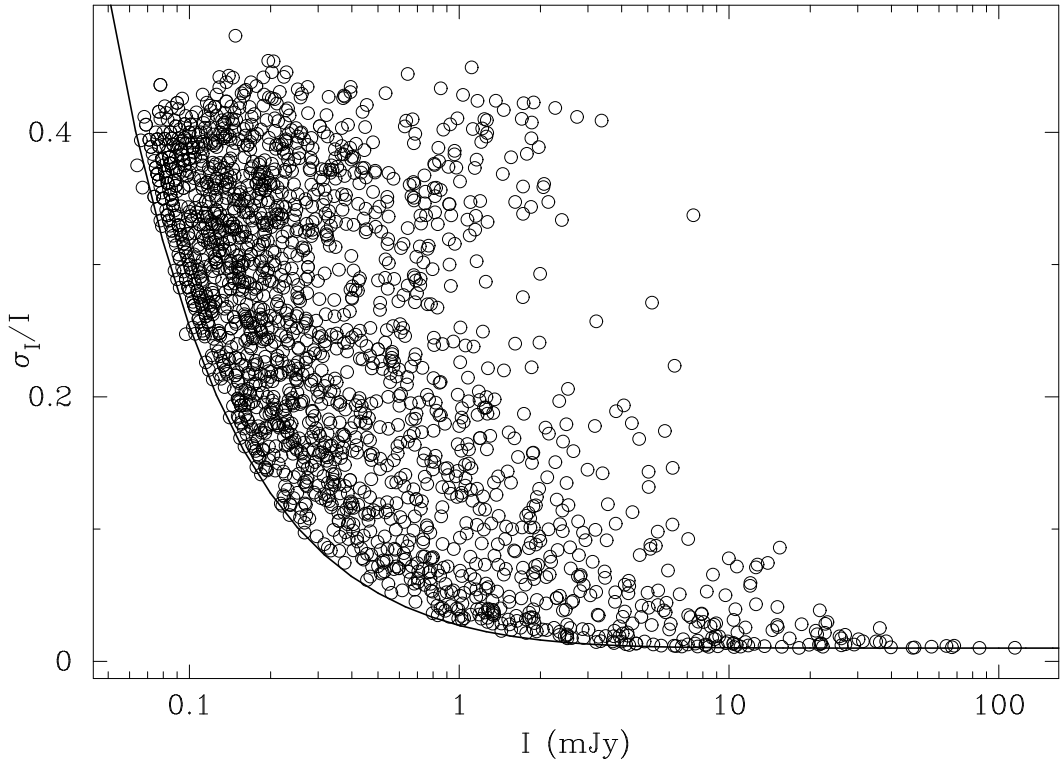}}
\caption{Relative flux density uncertainties as a function of integrated
flux density for the complete PDF catalogue. The solid line shows the
expected locus for point sources as given by Equation~\ref{pointerrs},
for a region of rms noise background of $16\,\mu$Jy. For bright sources, well
above the noise level, the second term in Equation~\ref{pointerrs} dominates.
  \label{fluxerrs}}
\end{figure*}

The source measurements, both position and flux density, seem
internally consistent within the expected uncertainties. To explore
the actual values of the uncertainties, {\sc sfind} was used to compile
a catalogue for the full PDF mosaic, again using an FDR threshold of 10\%,
giving a total of 2058 sources (see \S~\ref{catalogues}). The relative
uncertainties in the integrated flux density are shown in
Figure~\ref{fluxerrs}, with a line indicating the expected uncertainties for
point sources as given by Equation~\ref{pointerrs}, using a background rms
noise level of $16\,\mu$Jy. This assumed background level is only significant
for the fainter sources, since at high S/N the second term in
Equation~\ref{pointerrs} dominates. The value of the assumed background
level selected corresponds to the local rms noise level in the
vicinity of the faintest detected sources.

\section{Catalogue construction}
\label{catalogues}

The catalogue for the full PDF was compiled using {\sc sfind} with
a 10\% FDR threshold, resulting in 2090 sources. The source parameters
recorded in the catalogue are the source position, peak and integrated
flux densities and uncertainties, source size and orientation (major
and minor axis full width at half maxima, and position angle measured east
of north), and the rms noise level in the image at the location of the
source. The source sizes have not been deconvolved from the synthesised
beam size, so sources of $12''\times6''$ are point sources. Part of the
{\sc sfind} algorithm includes checking for source parameters indicating
an object smaller than the synthesised beam size. Following the suggestion
of \citet{Con:97}, such sources are automatically re-fit assuming they are
point sources, (i.e., size and position angle are taken to be those of the
synthesised beam, and only position and flux density are derived), and
retained if the fitting converges.

\begin{figure*}
\centerline{\includegraphics[angle=-90,width=15cm]{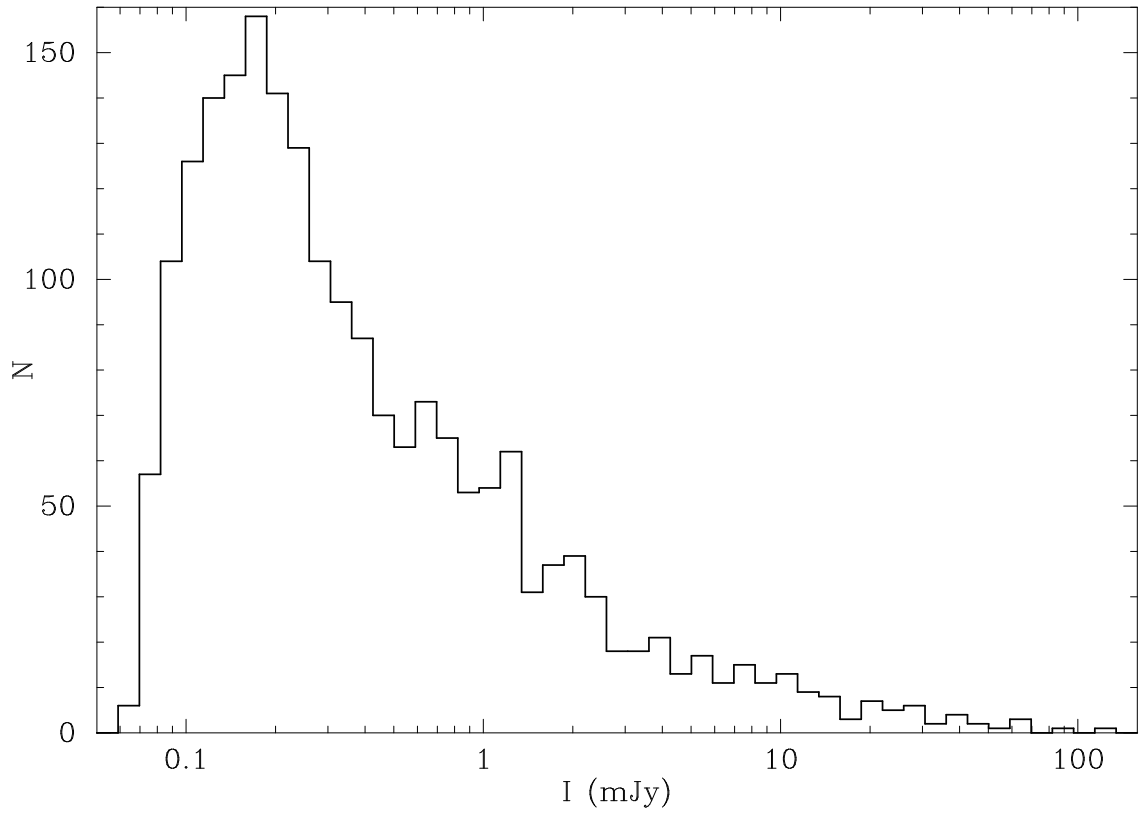}}
\caption{Histogram of integrated flux densities for the sources detected
in the primary PDF catalogue.
  \label{flxhist}}
\end{figure*}

A number of objects clearly associated with imaging artifacts, determined
through subsequent visual inspection of all the catalogued sources,
were removed, leaving 2058 sources in the final primary PDF catalogue.
Spurious objects will still remain in this catalogue, up to 10\% of
the total sources as given by the FDR threshold used \citep{Hop:02}.
Visual inspection of most of the few extended or complex sources in the
image confirm that many of these are poorly represented by the automated
detection process, usually being detected as several overlapping Gaussian
sources. We have not altered the detection parameters for these at all,
primarily since the total number of such objects is small and the few
poorly estimated parameters will not adversely affect the global properties
of the catalogue. Maintaining the easily quantified characteristics
of the catalogue construction is also important for understanding the
relative numbers of false detections, for example, while still ensuring
such complex sources remain in the catalogue in some form for subsequent
analysis. A histogram showing the complete distribution of integrated
flux densities is shown in Figure~\ref{flxhist}, and Table~\ref{cattab}
shows a short extract from the primary PDF catalogue. In Table~\ref{cattab}
and in the source count analysis below we adopt the notation $S_{\rm peak}$
and $S_{\rm int}$ rather than $S$ and $I$, since $S$ is commonly used
in diagrams of source counts to indicate the integrated flux density.
We follow that convention here in our subsequent Figures.

In addition to the primary PDF catalogue, we also independently compiled
a separate catalogue of sources in a region of $33'\times33'$
($1000\times1000$ pixels) centered on the most sensitive portion of the
survey. The actual flux density threshold derived for this region happens
to be slightly lower than that for the primary PDF catalogue, even though
a 10\% FDR threshold was specified for each. This emphasises
a somewhat non-intuitive feature of FDR thresholds, a result of their
adaptive nature. It turns out that simply because of the lower noise level,
the most sensitive region of the survey has a higher surface density of
detectable sources than the survey as a whole. So to maintain the
requested false discovery rate of 10\% a lower threshold is required
(giving more detections). The full survey area, however, even though
it contains this region within it, has a lower average surface density
of sources, and enough sources to reach a 10\% false detection rate are
detected at a higher threshold. This can also be understood by considering
the effects of differing source surface densities on diagrams like
Figure~1 of \citet{Hop:02}. The main result of this adaptive effect of
the FDR threshold for the current analysis is that the independent
catalogue from the deepest region contains fainter sources than the
primary source catalogue for the PDF, as well as a number of
other sources absent in the primary catalogue. This also explains why an
rms noise level of $16\,\mu$Jy is seen in Figure~\ref{fluxerrs} rather than
the lowest noise level in the image.

The effect of choosing different FDR thresholds (say 5\% or 1\%)
is, as expected, to reduce the numbers of faint sources detected, while
still detecting the same high S/N objects. This suggests that the majority
of the falsely-detected sources lie at flux densities close to the survey
limit, which is as expected for any thresholding technique. As discussed
further below, the primary effect of this will be to increase the
uncertainties on the estimate of the source counts at the faintest flux
densities. It is important to note that the actual fraction of falsely
detected sources is constrained to be {\em at most\/} the FDR fraction
specified (although the actual fraction cannot be known {\em a priori\/}),
and can be significantly less in some cases, as seen in the simulations of
\citet{Hop:02}. So the optimistic viewpoint would be that there is {\em less\/}
than a 10\% false detection rate in a 10\% FDR threshold catalogue. The effect
of choosing a very conservative threshold (1\%, say) is to increase the
reliability of the detected sources at the expense of missing many real,
fainter sources. Since a primary goal of the PDS is to fully exploit the
sensitivity of the deep 1.4\,GHz survey, the more liberal FDR threshold of
10\% was selected to maximise the number of sources close to the survey
threshold while maintaining an acceptable rate of false detections.

Alternative catalogues can (and should) also be constructed from the
PDF by using different thresholds (either FDR-determined or more traditionally
by specifying some $\sigma$-related threshold), or different algorithms.
The possibility of eventually seeing a robust, consistent source list
constructed from independently compiled catalogues is one reason we have
chosen to make the PDF image public.

\section{Source counts}
\label{scounts}

With the catalogues now available we proceed to construct the
1.4\,GHz differential source counts for the PDF, and to compare them to
earlier source count estimates. We proceed in a similar fashion to that
described in \citet{Hop:99b} and \cite{Hop:98b}. To account for
extended sources with integrated flux densities above the catalogue limits
but missed by the detection algorithm due to having peak flux densities below
the detection limit, a resolution correction was applied using the
same method and form as in \cite{Hop:98b}. A weighting correction is
also applied to account for the varying areas over which sources of
different flux densities could be detected. Detailed
descriptions of the weighting and resolution corrections, and their necessity,
are given elsewhere \citep{Hop:98b,Wind:84,Con:82,Oos:78,Kat:73}.

Since one primary goal of constructing the catalogues is to identify
a large population of radio-selected star-forming galaxies based on
complementary optical photometry and spectroscopy, allowing a certain fraction
of falsely detected sources to enter the catalogue is acceptable, as false
sources may be less likely to have optical counterparts (although this is
a strong function of the depth of the optical survey being compared with).
For the purposes of constructing source counts, though, the presence of
false sources may bias the resulting counts high. We have made no
attempt to remove such sources from the catalogue prior to constructing
the source counts, however, as it is not {\em a priori\/} known which
sources are false. Since it is likely that the majority of the false sources
will occur at the faintest flux density levels, where the uncertainties in
the resolution and weighting corrections will be largest, the most likely
outcome is that the uncertainty in the faintest few source count bins will
be a little larger than estimated.

In calculating the weighting correction, the effective area over which each
source could be detected is determined. The effective area for the whole
PDF is shown in Figure~\ref{areavsflx} as a function of the rms noise
level in the mosaic. The fraction of the total survey area is derived from
both the theoretical noise image, constructed using the {\sc miriad} task
{\sc linmos}, and as derived from measurements made in the mosaic image
itself. The theoretical rms noise level in the image reaches below
$10\,\mu$Jy, consistent with the earlier results of \citet{Hop:99b}. The
source detection, though, makes use of the measured image noise level, and
it is these values which need to be used in determining the effective
area over which each source could be detected. In constructing the
source counts for the deep independent region, the weighting corrections
derived for a given rms noise level are smaller than for the whole
PDF. This is because the deep independent region has a more uniform
noise level over a smaller total extent.

It should be noted that since we do not make any attempt to combine
the components of multiple component sources, the source counts
constructed from such a ``component catalogue" (rather than a ``source
catalogue") may be somewhat biased. The extent of this effect will be
flux density dependent and (as discussed in \S~\ref{disc} below)
of the order of about $10\%$ around 1\,mJy. With the majority of 
the PDF sources being fainter than 1\,mJy, and with the decreasing
contribution of multiple-component AGN type sources at sub-mJy levels,
the extent of any such bias in this regime is expected to be small.
Above about 1\,mJy, though, such an effect may contribute to the source
counts constructed herein.

A further issue regarding possible sources of uncertainty in constructing
the source counts is related to objects close to the detection limit
with large $S_{\rm int}/S_{\rm peak}$. These objects will have a large
weighting correction by virtue of only being detectable over a small
effective area. With a large integrated flux density that correction
translates into a disproportionate contribution to a flux density bin
where the majority of sources contribute fairly low weights. This can
result in a source count bin with both a comparatively high count
and a large uncertainty. If many bins are affected by objects like
this the source counts can appear both biased upwards and highly
variable over a small range in flux density. Investigation of the
measured sizes of low S/N objects ($S_{\rm peak}/\sigma<5$) reveals
that they have larger median sizes than higher S/N sources at
corresponding flux densities, suggesting that their sizes and
hence integrated flux densities are overestimated. To avoid such
biases from these sources, objects in the primary PDF catalogue with
$S_{\rm int}/S_{\rm peak}>2.5$ and $S_{\rm peak}/\sigma<5$ are deemed
to contribute to the flux density bin corresponding to their peak flux
density (as though they were point sources) when constructing the source
counts, rather than that for their overestimated integrated flux density.
Similarly, for the deep independent catalogue with a different threshold
level, limits of $S_{\rm int}/S_{\rm peak}>1.6$ and $S_{\rm peak}/\sigma<5.5$
were established to eliminate these artifacts.
The number of such sources is 73 out of 2058 for the primary catalogue and
71 out of 491 for the deep independent catalogue (where a larger fraction
of sources are closer to the survey limit), so there should be little
effect on the source counts themselves apart from avoiding this type
of artifact. It was found that omitting these sources entirely (on
the tentative assumption that such objects might be false sources)
had the effect only of marginally lowering the derived source counts in
the faintest flux density bin for each catalogue. They have been retained
in the source counts presented here.

With this concern addressed, we now construct a resolution correction
($r$), as given in \citet{Hop:98b}, and a weighting correction
($w=T/D$, where $T$ is the total survey area and $D$ is the effective
area over which the source could be detected) for each source,
based on that source's integrated and peak flux densities respectively.
Each source thus contributes a number of effective sources, $N_{\rm eff}=rw$,
to the flux density bin in the source counts in which its integrated flux
density falls. Uncertainties on this value in each source count bin are the
rms counting errors, $(\sum_i (r_iw_i)^2)^{0.5}$. The width of the flux
density bins are chosen to include a minimum number of actual sources per
bin, apart from the brightest bin. The minimum number chosen was 150 sources
per bin for the primary catalogue for whole mosaic image, and 60 for the
deep independent catalogue. Ensuring such large numbers of actual
sources are present in each bin should minimise fluctuations in the
resulting source counts. The final source counts are constructed by
dividing these total effective numbers of sources per bin by the total
survey area and the bin width, to give the number of sources per unit
area per unit flux density, and then normalising by the Euclidean slope
of $S^{-2.5}$. The reason for the normalisation step is historical in
origin, relating to the early cosmological goals of studying
radio source counts, and the greater ease of distinguishing different forms
of evolution after such normalisation was made \citep{Ryle:68,Lon:66}.
It has persisted in the literature of radio source count investigations,
and is similarly used here for ease of comparison with other surveys.
We do not, however, normalise by the additional factor of 225 that is
sometimes used to fix the value of the differential source count
to unity at a flux density of 1\,Jy.

\section{Results}
\label{results}

The differential source counts were calculated for both the primary PDF
catalogue and the deep independent catalogue, and the results are presented
in Tables~\ref{srccnttab1} and \ref{srccnttab2} and are shown in
Figure~\ref{srccnt1}. Tables~\ref{srccnttab1} and \ref{srccnttab2} give the
flux density extrema and the mean flux density of each bin, the actual number
of sources in each bin (N), and the effective number of sources
(N$_{\rm eff}$) over the whole 4.56 square degree area after applying the
weighting and resolution corrections, along with the resulting source counts.
The mean flux density in each bin was calculated from the flux densities
of the detected sources in each bin, except towards the brighter flux densities
where the formula given by \citet[their equation 19]{Wind:84} was used,
making use of the known slope of the source counts to more accurately estimate
the mean flux density in a bin. This is especially important for the brightest
bin from each catalogue, where fewer objects are available to accurately
represent the mean flux density of a bin.
Figure~\ref{srccnt1} indicates the flux density range for each bin
by a horizontal bar, while the vertical uncertainties are the
statistical uncertainties in the counts, described above.
The primary and deep PDF catalogues allow the source counts constructed
from the PDF to span a flux density range from about $0.05-100$\,mJy,
although the sampling becomes sparse at brighter flux densities.
This reflects the limited area of the survey, which sharply restricts
the number of sources detectable at brighter flux densities. In
the counts from the deep independent catalogue, covering a smaller
area still, this effect is seen to appear at lower flux densities.

Given that the resolution and weighting corrections have been carefully
applied, the drop off in the source counts seen towards the faintest
flux densities should not immediately be discounted as incompleteness
in the counts. Also, since the faintest few source count bins may be
biased high by the presence of falsely detected sources, it is possible
that the observed drop off may be even greater. The angular size distribution
from which the resolution correction is derived, however, has larger
uncertainties at these lower flux densities, which could be compounded by
the actual incompleteness present in the source catalogues. This may have
the effect of producing the observed drop off in the measured source counts
at these low flux densities. This bias would be in the opposite direction to
that expected from the presence of falsely detected sources. In addition,
the source counts are eventually expected to converge below a few $\mu$Jy,
in order for their integrated sky brightness not to distort the observed
cosmic background radiation spectrum at centimeter wavelengths
\citep{Wind:93}. Although it is unlikely that this effect is occurring
at the flux densities probed here, the observed source counts may be
starting to give an indication of a new slope change at
levels around $50-100\,\mu$Jy \citep[but c.f.][]{Rich:00}.

Figure~\ref{srccnt2} shows the newly derived PDF source counts and adds
other source count estimates from existing surveys for comparison.
The compilation of source counts from \citet{Wind:93} is shown, as well
as those from the Faint Images of the Radio Sky at Twenty centimeters
survey \citep[FIRST,][]{Whi:97}, the counts from the Very Large Array
(VLA) observations of the Hubble Deep Field \citep[HDF,][]{Rich:00},
and those of \citet{Pran:01}.
A previous estimate of the source counts from an earlier ATCA mosaic
of the Phoenix region \citep{Hop:99b} is also shown for comparison.
The similarities and discrepancies between these various results are
discussed in \S~\ref{disc} below.

The solid line in both Figures~\ref{srccnt1} and \ref{srccnt2} is a
linear least squares sixth order polynomial fit, derived in order to aid
other workers in this field. The fit was constructed using the newly
derived PDF counts from both the deep and primary catalogues. Given the
sparsity of our sampling of the source counts above about 2\,mJy, and
our consistency with the counts from FIRST in that regime, we supplement
our counts with the FIRST source count values above 2.5\,mJy, where they
are complete. We also excluded the brightest PDF source count bin from
the fit since, although its uncertainties indicate it is consistent
with the counts from FIRST, it clearly lies lower than those counts
and if retained would bias the resulting fit somewhat.
The resulting polynomial fit is given by
\begin{equation}
\log[(dN/dS)/(S^{-2.5})] = \sum_{i=0}^{6} a_i(\log[S/{\rm mJy}])^i,
\end{equation}
with $a_0=0.859$, $a_1=0.508$, $a_2=0.376$, $a_3=-0.049$, $a_4=-0.121$,
$a_5=0.057$ and $a_6=-0.008$. This fit is valid over the flux density
range $0.05-1000\,$mJy. A previously published fit to the source counts
down to about $100\,\mu$Jy was of a third order polynomial by
\citet{Kat:88}, also shown in Figure~\ref{srccnt2} for
comparison. Our sixth order polynomial is
substantially similar to this third order one above $100\,\mu$Jy.
A higher order polynomial is necessary to account for both the number
of changes of slope in the observed counts, as well as the sense of
the curvature below $100\,\mu$Jy. A fourth order polynomial would be
sufficient to model these features, but it was found that in this case the
point of inflection around 0.5\,mJy was poorly represented. To ensure that
the polynomial showed concave down curvatures at the extremes, a sixth
order polynomial was thus required. The residuals from the sixth
order fit above have an rms of about 0.04 in the logarithm of
the normalised counts.

\section{Discussion}
\label{disc}

While it is clear that there is a high level of consistency between
most source count estimates over a broad range of flux densities,
and the PDF source counts are consistent with the compilation of
source counts given by \citet{Wind:93} down to about $150\,\mu$Jy,
there are discrepancies which warrant attention. These are primarily the
apparent inconsistency with earlier PDF counts
\citep{Hop:99b}, the discrepancy around 1\,mJy between several
catalogues, and the discrepancy with the deep VLA counts of the HDF 
\citep{Rich:00}. We explore these discrepancies here.

We have established that the earlier PDF source count estimate suffers from
the artifact due to overestimates of source extent, and hence integrated
flux density, for objects detected near the threshold of the survey as
described in \S~\ref{scounts} above. This results in an overestimate in the
counts from about $150-300\,\mu$Jy and an underestimate below about
$100\,\mu$Jy. Also, the strong variation from point to point
in these counts is attributable to the relatively low minimum number of
objects (30) per bin used in generating those counts. When these issues are
addressed the counts derived from that earlier catalogue are seen to be
identical with the current estimate over the full flux density range probed.

At flux densities below about 2\,mJy the source counts from FIRST
systematically drop away from the compilation of counts from \citet{Wind:93}.
This is described by \citet{Whi:97} as due to the low peak flux densities
of many faint extended sources undetectable in the FIRST survey, and
this feature is thus an indicator of the incompleteness of the survey
at its faint limit. As mentioned above, in determining the polynomial fit 
we have used only FIRST source count points from flux density bins
brighter than 2.5\,mJy to avoid these incomplete bins.

The source counts from the Australia Telescope ESO Slice Project (ATESP)
survey \citep{Pran:01} around 1\,mJy are also lower than the PDF
counts and the source count compilation. The ATESP survey covers a large
area (26 square degrees) with a fairly uniform rms noise level
(fluctuating between about $70-100\,\mu$Jy), and detects sources down
to about 0.5\,mJy \citep{Pran:00a,Pran:00b}. The discrepancy is seen in
the two ATESP bins spanning $0.7-1.4\,$mJy, so the inconsistency
between these counts and those from the PDF occurs at a relatively high S/N
level ($\sim10\sigma$). Recognising that great effort has been spent in
ensuring that the catalogues and source counts for both ATESP and the PDF are
as reliable as possible, we have explored this issue in some detail in an
attempt to reconcile the two.
Using the publicly available ATESP catalogue and the primary PDF catalogue
we have directly compared actual source surface densities, and we find that
the difference is present in the catalogues, rather than being an artifact
of the construction of the source counts. The surface density of sources
between the catalogues was found to differ by the same fraction as the
source counts.

The ATESP team have already established that field to field variance
resulting from galaxy clustering could account for perhaps roughly half
this discrepancy. They constructed source counts for independent regions
within their large survey area to explore the extent of the fluctuations
seen. From the angular correlation functions of \citet{Age:00b} and
\citet{Wind:90}, such effects are likely to be on the order of
$10-20\%$ of the measured source counts, and this is consistent with
the level of variation seen within the ATESP study. This is not
sufficient to account for the full discrepancy seen around 1\,mJy.
We then explored differences between the source detection procedures
used (ATESP uses the task {\sc imsad} in {\sc miriad} for source detection).
By using {\sc sfind} to measure sources in several of the individual
ATESP mosaics, we established the following results. The {\sc sfind}
and {\sc imsad} measurements are highly consistent, within expected
Gaussian fitting uncertainties, for the majority of sources. At very
low S/N there are some inconsistencies between the two tasks in the
measurements for a small fraction of sources, related to the selection
of pixels for inclusion in the Gaussian fitting, but even for these
objects the measured integrated flux densities are mostly consistent.
The surface densities of the {\sc sfind} catalogues for the ATESP mosaics
are consistent with that of the PDF catalogue, although in the range of the
discrepancy the PDF catalogue is at the upper end of the observed field
to field variations between the ATESP mosaics. The main difference between
the {\sc imsad} and {\sc sfind} catalogues in the ATESP mosaics appears
to be explained by complex sources split into components during the
fitting procedure by both tasks. The ATESP survey has examined and refit
such complex sources individually to construct a ``source catalogue" rather
than a ``component catalogue," whereas the present investigation for the
Phoenix survey neglects this step for reasons given in \S\ref{catalogues}
above. When the ATESP source counts are constructed using the
``component catalogue" the source count value is increased by about
$10\%$ in the flux density region of the discrepancy
(Prandoni 2002, private communication).

From these results we conclude that the discrepancy between the ATESP
and PDF source counts is attributable to two main effects. First, calculating
the source counts based on a ``component catalogue" rather than a
``source catalogue" increases the resulting source count by about $10\%$.
This combined with a field to field variation putting the PDF counts to
the higher end of observed variations in this flux density range,
consistent with expected variations from measurements of the angular
correlation function, is sufficient to account for the observed
discrepancy.

The deep VLA source counts of the HDF region, compiled by \citet{Rich:00},
also show a discrepancy with the PDF source counts and those in
the source count compilation. Above about $100\,\mu$Jy the HDF
counts deviate by up to a factor of two below the counts of other
surveys. This discrepancy has already been described in detail
by \citet{Rich:00}, and we can add little to that discussion
other than the comment that the extent of this deviation again seems to
be larger than can be explained through the expected fluctuations from
galaxy clustering \citep{Age:00b,Wind:90}.

Apart from these discrepancies, the 1.4\,GHz source counts seem now to
be consistently determined from the brightest levels down to about
$50\,\mu$Jy. The sixth order polynomial fit to the counts given
here will provide a useful parametrisation of the counts above this level.

\section{Summary}
\label{summ}

We have used the ATCA to construct a 1.4\,GHz mosaic image of slightly more
than 4.5 square degrees. This image reaches rms noise levels of about
$12\,\mu$Jy at its most sensitive, and a primary catalogue of 2058 radio
sources in the mosaic has been constructed with a 10\% false discovery rate.
Detailed analysis of the source measurement in independently observed
pointings confirm the reliability of the catalogued source parameters,
consistent with the expected uncertainties from the image noise and the
Gaussian source fitting. Differential source counts were constructed for
both a deep independent catalogue of a $33'\times33'$ region centered on
the most sensitive portion of the survey, and the full primary PDF catalogue.
The source counts are seen to be consistent with previous surveys of
similar size, while sampling to fainter flux densities, and sensitivity,
while sampling a larger area.
Discrepancies between the ATESP source counts and those of the PDF
at levels of 1\,mJy have been explored in detail. These are attributable
to a field to field variation increasing the PDF counts slightly over the flux
density range of the discrepancy, and are otherwise consistent with
the source counts derived from the ATESP ``component catalogue."
We have performed a sixth order polynomial fit to our derived source
counts, supplemented at the bright end by those from FIRST, which
adequately parametrises the source counts from $0.05-1000\,$mJy.
The 1.4\,GHz radio mosaic and catalogues are available on the web at
\verb+http://www.atnf.csiro.au/~ahopkins/phoenix/+
or from the authors by request.

\acknowledgements

The authors would like to thank the referee for several very helpful
suggestions, and extend their warmest thanks to Isabella Prandoni for helpful
discussions and for kindly making several of the ATESP mosaics available for
analysis.
AMH gratefully acknowledges support provided by NASA through Hubble
Fellowship grant HST-HF-01140.01-A awarded by the Space Telescope Science
Institute, which is operated by the Association of Universities for
Research in Astronomy, Inc., for NASA, under contract NAS 5-26555.
JA gratefully acknowledges the support from the Science and Technology
Foundation (FCT, Portugal) through the fellowship BPD-5535-2001 and the
research grant ESO-FNU-43805-2001.
The Australia Telescope is funded by the Commonwealth of Australia
for operation as a National Facility managed by CSIRO.

\begin{deluxetable}{cccc}
\tablewidth{0pt}
\tablecaption{Mosaic pointings observed with the ATCA.
 \label{points}}
\tablehead{
\multicolumn{2}{c}{1994} & \multicolumn{2}{c}{$1997-2001$}\\
\colhead{RA (J2000)} & \colhead{Dec (J2000)} & \colhead{RA (J2000)} & \colhead{Dec (J2000)}
}
\startdata
$01~09~13.589$ & $-45~44~01.41$ & $01~08~44.200$ & $-45~30~00.00$ \\
$01~10~13.123$ & $-45~59~02.70$ & $01~08~44.200$ & $-45~45~00.00$ \\
$01~10~13.480$ & $-45~29~02.70$ & $01~08~44.200$ & $-46~00~00.00$ \\
$01~11~12.651$ & $-46~14~04.00$ & $01~09~58.600$ & $-45~22~30.00$ \\
$01~11~13.016$ & $-45~44~04.01$ & $01~09~58.600$ & $-45~37~30.00$ \\
$01~11~13.374$ & $-45~14~04.01$ & $01~09~58.600$ & $-45~52~30.00$ \\
$01~12~12.173$ & $-46~29~05.32$ & $01~09~58.600$ & $-46~07~30.00$ \\
$01~12~12.546$ & $-45~59~05.33$ & $01~11~13.000$ & $-45~15~00.00$ \\
$01~12~12.912$ & $-45~29~05.33$ & $01~11~13.000$ & $-45~30~00.00$ \\
$01~12~13.272$ & $-44~59~05.34$ & $01~11~13.000$ & $-45~45~00.00$ \\
$01~13~12.070$ & $-46~14~06.66$ & $01~11~13.000$ & $-46~00~00.00$ \\
$01~13~12.444$ & $-45~44~06.67$ & $01~11~13.000$ & $-46~15~00.00$ \\
$01~13~12.812$ & $-45~14~06.68$ & $01~12~27.400$ & $-45~22~30.00$ \\
$01~14~11.587$ & $-46~29~08.02$ & $01~12~27.400$ & $-45~37~30.00$ \\
$01~14~11.970$ & $-45~59~08.03$ & $01~12~27.400$ & $-45~52~30.00$ \\
$01~14~12.346$ & $-45~29~08.03$ & $01~12~27.400$ & $-46~07~30.00$ \\
$01~14~12.716$ & $-44~59~08.04$ & $01~13~41.800$ & $-45~30~00.00$ \\
$01~15~11.490$ & $-46~14~09.40$ & $01~13~41.800$ & $-45~45~00.00$ \\
$01~15~11.874$ & $-45~44~09.41$ & $01~13~41.800$ & $-46~00~00.00$ \\
$01~15~12.252$ & $-45~14~09.41$ & & \\
$01~16~11.003$ & $-46~29~10.79$ & & \\
$01~16~11.396$ & $-45~59~10.80$ & & \\
$01~16~11.782$ & $-45~29~10.80$ & & \\
$01~16~12.162$ & $-44~59~10.81$ & & \\
$01~17~10.911$ & $-46~14~12.21$ & & \\
$01~17~11.306$ & $-45~44~12.21$ & & \\
$01~17~11.694$ & $-45~14~12.22$ & & \\
$01~18~10.824$ & $-45~59~13.64$ & & \\
$01~18~11.220$ & $-45~29~13.65$ & & \\
$01~19~10.739$ & $-45~44~15.09$ & & \\
\enddata
\end{deluxetable}

\begin{deluxetable}{cccccccc}
\tablewidth{0pt}
\tablecaption{Extract from PDF catalogue.
 \label{cattab}}
\tablehead{
\colhead{RA} & \colhead{Dec} & \colhead{$S_{\rm peak}$} & \colhead{$S_{\rm int}$} & \colhead{$\theta_{\rm major}$} & \colhead{$\theta_{\rm minor}$} & \colhead{PA} & \colhead{$\sigma$} \\
\colhead{(J2000)} & \colhead{(J2000)} & \colhead{(mJy)} & \colhead{(mJy)} & \colhead{$('')$} & \colhead{$('')$} & \colhead{$(\,^{\circ})$} & \colhead{(mJy)}
}
\startdata
$1~08~26.641$ & $-45~42~27.38$  &  $0.524 \pm 0.029$  &  $0.524 \pm 0.040$  &  $12.0$  &  $6.0$  &  $0.0$  &  $0.020$ \\
$1~08~26.778$ & $-45~30~03.85$  &  $0.158 \pm 0.041$  &  $0.158 \pm 0.058$  &  $12.0$  &  $6.0$  &  $0.0$  &  $0.029$ \\
$1~08~27.499$ & $-46~07~42.57$  &  $0.433 \pm 0.036$  &  $0.543 \pm 0.062$  &  $13.1$  &  $6.9$  &  $2.9$  &  $0.025$ \\
$1~08~27.760$ & $-45~54~11.86$  &  $0.103 \pm 0.028$  &  $0.155 \pm 0.055$  &  $13.4$  &  $8.1$  &  $2.5$  &  $0.019$ \\
$1~08~27.979$ & $-45~43~24.75$  &  $0.330 \pm 0.028$  &  $0.330 \pm 0.040$  &  $12.0$  &  $6.0$  &  $0.0$  &  $0.020$ \\
$1~08~28.928$ & $-46~15~45.24$  &  $0.171 \pm 0.048$  &  $0.171 \pm 0.068$  &  $12.0$  &  $6.0$  &  $0.0$  &  $0.034$ \\
$1~08~28.932$ & $-45~40~21.55$  &  $1.122 \pm 0.031$  &  $1.343 \pm 0.049$  &  $13.1$  &  $6.6$  &  $5.1$  &  $0.020$ \\
$1~08~29.137$ & $-45~32~39.39$  &  $0.368 \pm 0.041$  &  $0.368 \pm 0.058$  &  $12.0$  &  $6.0$  &  $0.0$  &  $0.029$ \\
$1~08~29.367$ & $-46~11~43.36$  &  $0.155 \pm 0.045$  &  $0.222 \pm 0.086$  &  $14.9$  &  $6.9$  &  $8.5$  &  $0.031$ \\
$1~08~29.467$ & $-45~48~48.49$  &  $0.224 \pm 0.027$  &  $0.224 \pm 0.038$  &  $12.0$  &  $6.0$  &  $0.0$  &  $0.019$ \\
$1~08~30.033$ & $-45~11~37.96$  &  $0.284 \pm 0.054$  &  $0.428 \pm 0.107$  &  $15.1$  &  $7.2$  &  $-4.9$  &  $0.037$ \\
$1~08~30.244$ & $-45~39~52.15$  &  $0.140 \pm 0.031$  &  $0.251 \pm 0.070$  &  $16.5$  &  $7.8$  &  $11.2$  &  $0.021$ \\
$1~08~30.249$ & $-45~13~20.37$  &  $0.599 \pm 0.053$  &  $0.866 \pm 0.101$  &  $13.9$  &  $7.5$  &  $-3.1$  &  $0.036$ \\
$1~08~30.433$ & $-45~58~46.39$  &  $0.114 \pm 0.029$  &  $0.185 \pm 0.062$  &  $14.2$  &  $8.2$  &  $8.2$  &  $0.020$ \\
$1~08~30.586$ & $-45~48~01.06$  &  $0.164 \pm 0.027$  &  $0.164 \pm 0.038$  &  $12.0$  &  $6.0$  &  $0.0$  &  $0.019$ \\
$1~08~30.694$ & $-45~44~48.11$  &  $0.169 \pm 0.028$  &  $0.169 \pm 0.040$  &  $12.0$  &  $6.0$  &  $0.0$  &  $0.020$ \\
$1~08~30.711$ & $-46~03~47.59$  &  $0.257 \pm 0.031$  &  $0.351 \pm 0.056$  &  $12.6$  &  $7.8$  &  $-0.6$  &  $0.021$ \\
$1~08~30.719$ & $-46~04~04.88$  &  $0.124 \pm 0.030$  &  $0.124 \pm 0.042$  &  $12.0$  &  $6.0$  &  $0.0$  &  $0.021$ \\
$1~08~30.753$ & $-45~35~49.28$  &  $9.502 \pm 0.101$  &  $11.038 \pm 0.123$  &  $12.7$  &  $6.6$  &  $-0.7$  &  $0.024$ \\
$1~08~32.847$ & $-46~13~11.18$  &  $0.168 \pm 0.044$  &  $0.168 \pm 0.062$  &  $12.0$  &  $6.0$  &  $0.0$  &  $0.031$ \\
$1~08~33.018$ & $-45~52~40.20$  &  $0.599 \pm 0.028$  &  $0.782 \pm 0.049$  &  $13.0$  &  $7.3$  &  $-1.8$  &  $0.019$ \\
$1~08~33.159$ & $-45~45~43.37$  &  $0.223 \pm 0.029$  &  $0.288 \pm 0.050$  &  $12.1$  &  $7.7$  &  $10.1$  &  $0.020$ \\
$1~08~33.247$ & $-45~06~21.28$  &  $4.351 \pm 0.128$  &  $5.731 \pm 0.221$  &  $13.5$  &  $7.0$  &  $-5.7$  &  $0.083$ \\
$1~08~34.528$ & $-45~58~02.12$  &  $0.310 \pm 0.028$  &  $0.397 \pm 0.048$  &  $13.2$  &  $7.0$  &  $1.4$  &  $0.019$ \\
$1~08~34.541$ & $-46~06~03.47$  &  $0.660 \pm 0.029$  &  $0.660 \pm 0.041$  &  $12.0$  &  $6.0$  &  $0.0$  &  $0.020$ \\
$1~08~34.788$ & $-46~16~18.63$  &  $0.299 \pm 0.043$  &  $0.431 \pm 0.081$  &  $14.1$  &  $7.3$  &  $-3.3$  &  $0.029$ \\
$1~08~34.846$ & $-45~54~46.14$  &  $0.100 \pm 0.025$  &  $0.207 \pm 0.064$  &  $15.5$  &  $9.6$  &  $23.8$  &  $0.017$ \\
$1~08~35.179$ & $-45~16~46.28$  &  $0.213 \pm 0.041$  &  $0.213 \pm 0.058$  &  $12.0$  &  $6.0$  &  $0.0$  &  $0.029$ \\
$1~08~35.301$ & $-46~15~01.38$  &  $0.637 \pm 0.043$  &  $0.867 \pm 0.077$  &  $14.0$  &  $7.0$  &  $-0.2$  &  $0.029$ \\
$1~08~35.959$ & $-45~46~59.15$  &  $0.116 \pm 0.026$  &  $0.157 \pm 0.047$  &  $12.3$  &  $8.0$  &  $7.2$  &  $0.018$ \\
\enddata
\end{deluxetable}

\begin{deluxetable}{ccccc}
\tablewidth{0pt}
\tablecaption{Source counts from deepest PDF region.
 \label{srccnttab1}}
\tablehead{
\colhead{Range in $S_{1.4}$} & \colhead{$\left<S_{1.4}\right>$} & \colhead{N} & \colhead{N$_{\rm eff}$} & \colhead{(dN/dS)$/S^{-2.5}$} \\
\colhead{(mJy)} & \colhead{(mJy)} & & & \colhead{(Jy$^{1.5}$\,sr$^{-1}$)}
}
\startdata
$0.048-0.067$  &  $0.057$  &  64  &   149.9  &  $2.49 \pm 0.49$ \\
$0.067-0.078$  &  $0.072$  &  64  &    75.2  &  $3.27 \pm 0.41$ \\
$0.078-0.095$  &  $0.086$  &  66  &    86.0  &  $3.66 \pm 0.58$ \\
$0.095-0.115$  &  $0.104$  &  63  &    70.0  &  $4.00 \pm 0.50$ \\
$0.115-0.150$  &  $0.131$  &  61  &    65.8  &  $3.99 \pm 0.51$ \\
$0.150-0.202$  &  $0.174$  &  60  &    63.6  &  $5.04 \pm 0.65$ \\
$0.202-0.409$  &  $0.288$  &  60  &    62.7  &  $4.56 \pm 0.59$ \\
$0.409-18.2$  &  $2.75$  &  53  &    54.8  &  $13.04 \pm 1.79$ \\
\enddata
\end{deluxetable}

\begin{deluxetable}{ccccc}
\tablewidth{0pt}
\tablecaption{Source counts from primary PDF catalogue.
 \label{srccnttab2}}
\tablehead{
\colhead{Range in $S_{1.4}$} & \colhead{$\left<S_{1.4}\right>$} & \colhead{N} & \colhead{N$_{\rm eff}$} & \colhead{(dN/dS)$/S^{-2.5}$} \\
\colhead{(mJy)} & \colhead{(mJy)} & & & \colhead{(Jy$^{1.5}$\,sr$^{-1}$)}
}
\startdata
$0.089-0.111$  &  $0.100$  &  159  &  1172.2  &  $3.91 \pm 1.25$ \\
$0.111-0.133$  &  $0.122$  &  157  &   804.7  &  $4.25 \pm 0.90$ \\
$0.133-0.159$  &  $0.146$  &  154  &   606.9  &  $4.44 \pm 0.41$ \\
$0.159-0.186$  &  $0.172$  &  157  &   492.8  &  $4.96 \pm 0.48$ \\
$0.186-0.227$  &  $0.206$  &  157  &   437.4  &  $4.73 \pm 0.39$ \\
$0.227-0.282$  &  $0.253$  &  150  &   353.7  &  $4.67 \pm 0.39$ \\
$0.282-0.376$  &  $0.326$  &  155  &   297.9  &  $4.37 \pm 0.36$ \\
$0.376-0.531$  &  $0.447$  &  151  &   236.6  &  $4.64 \pm 0.39$ \\
$0.531-0.759$  &  $0.635$  &  150  &   191.5  &  $6.15 \pm 0.52$ \\
$0.759-1.18$  &  $0.945$  &  150  &   168.2  &  $7.98 \pm 0.66$ \\
$1.18-2.32$  &  $1.65$  &  153  &   164.7  &  $11.50 \pm 0.94$ \\
$2.32-10.5$  &  $4.91$  &  152  &   160.8  &  $23.98 \pm 1.96$ \\
$10.5-115$  &  $33.3$  &  57  &    60.0  &  $84.39 \pm 11.18$ \\
\enddata
\end{deluxetable}

\begin{figure*}
\centerline{\includegraphics[angle=-90,width=15cm]{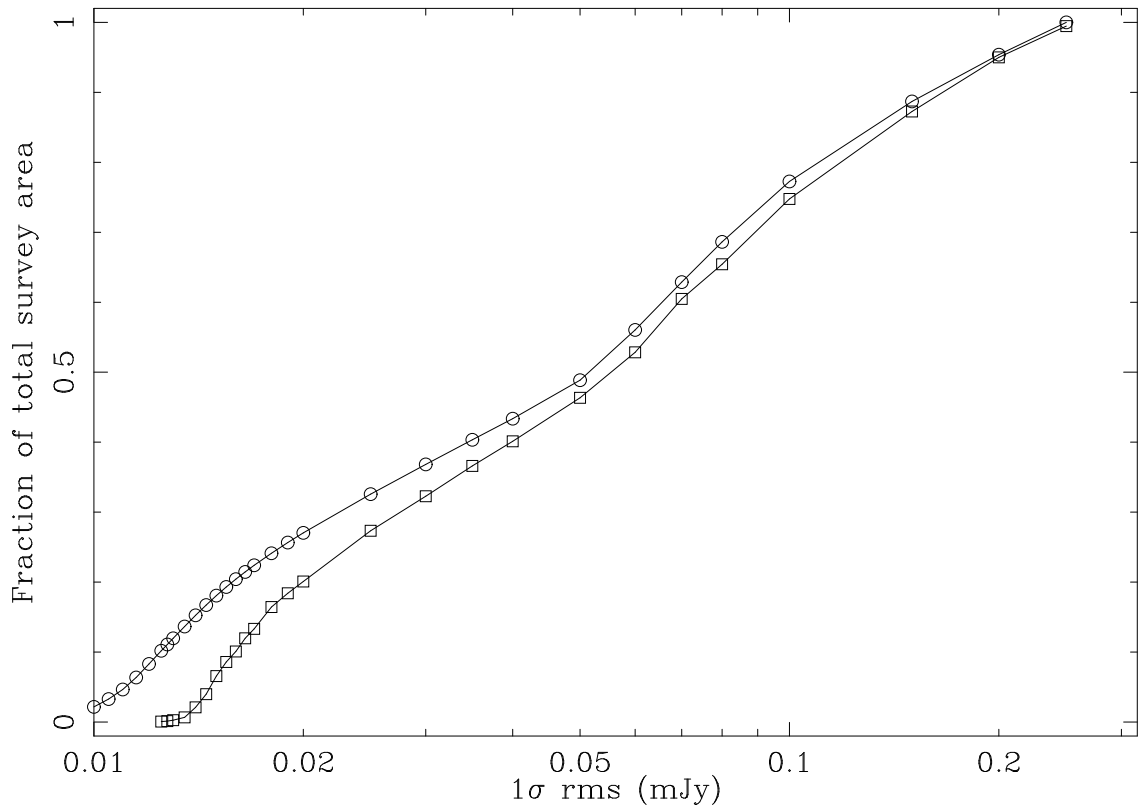}}
\caption{The fraction of the survey area having an rms noise level equal to or
lower than the flux densities indicated, used in constructing the
weighting corrections for the source counts. The circles represent the
areas derived from the theoretical noise level image for the mosaic, the
squares give the actual measured values.
  \label{areavsflx}}
\end{figure*}

\begin{figure*}
\centerline{\includegraphics[angle=-90,width=14.5cm]{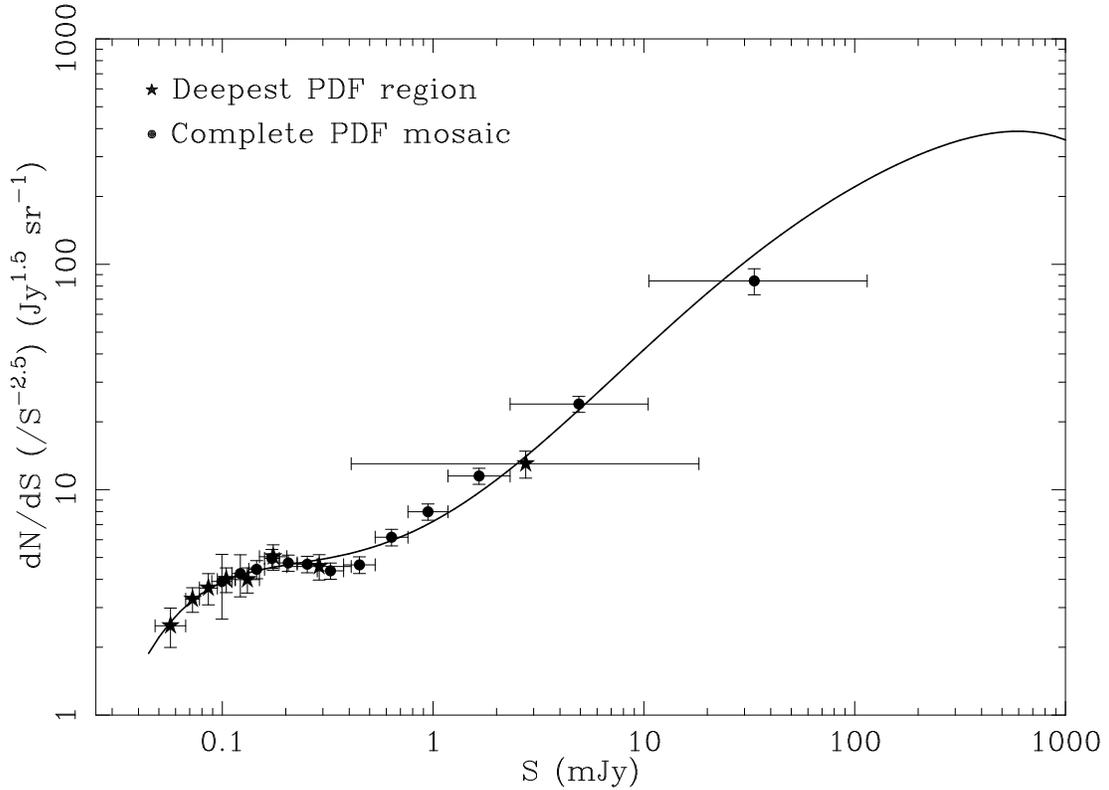}}
\caption{1.4\,GHz differential source counts for the deep independent
$33'\times33'$ region within the PDF (stars), and the primary catalogue
(circles). Horizontal uncertainties represent the flux density range
of each source count bin. The line is a sixth order polynomial fit
described in the text.
  \label{srccnt1}}
\end{figure*}

\begin{figure*}
\centerline{\includegraphics[angle=-90,width=14.5cm]{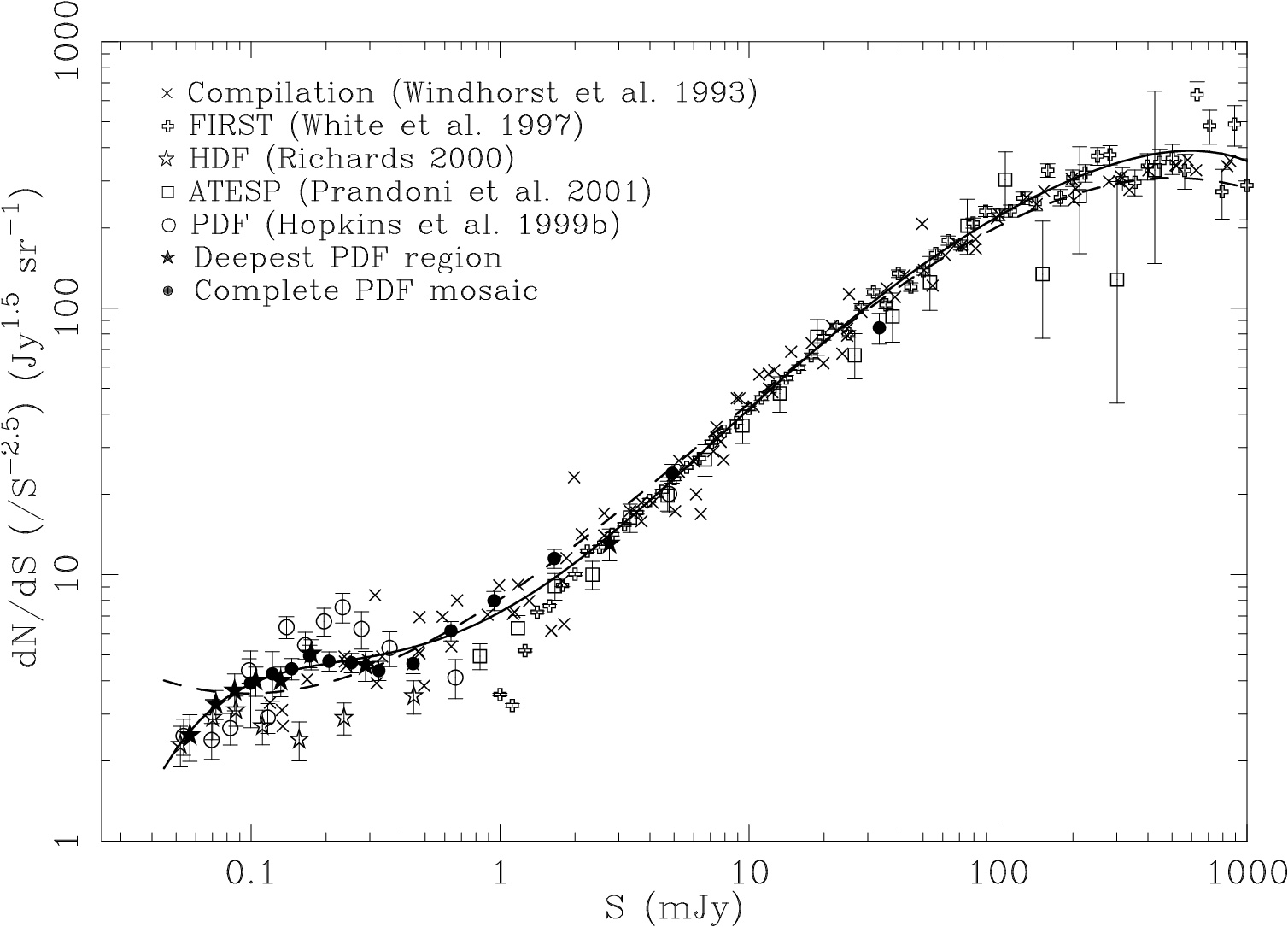}}
\caption{1.4\,GHz differential source counts for the primary PDF catalogue
(solid circles) and deep independent catalogue (solid stars).
For comparison we also show the compilation of source counts from
\citet{Wind:93}, source counts from the FIRST survey \citep{Whi:97},
those from the Australia Telescope ESO Slice Project (ATESP) survey
\citep{Pran:01}, and from VLA observations
of the HDF \citep{Rich:00}. Previously estimated source counts from an
earlier ATCA mosaic of the Phoenix area are also shown \citep{Hop:99b}. 
The solid line is a sixth order least squares polynomial fit. The third
order polynomial fit from \citet{Kat:88} is also shown for comparison
(dashed line).
  \label{srccnt2}}
\end{figure*}

\end{document}